# SYNTHESIZING STUDY-SPECIFIC CONTROLS USING GENERATIVE MODELS ON OPEN ACCESS DATASETS FOR HARMONIZED MULTI-STUDY ANALYSES

*Shruti P. Gadewar, Alyssa H. Zhu, Iyad Ba Gari, Sunanda Somu, Sophia I. Thomopoulos,*
*Paul M. Thompson, Talia M. Nir, Neda Jahanshad*

Imaging Genetics Center, Mark and Mary Stevens Neuroimaging and Informatics Institute,
Keck School of Medicine, University of Southern California, Marina del Rey, CA, USA

## ABSTRACT

Neuroimaging consortia can enhance reliability and generalizability of findings by pooling data across studies to achieve larger sample sizes. To adjust for site and MRI protocol effects, imaging datasets are often harmonized based on healthy controls. When data from a control group were not collected, statistical harmonization options are limited as patient characteristics and acquisition-related variables may be confounded. Here, in a multi-study neuroimaging analysis of Alzheimer's patients and controls, we tested whether it is possible to generate synthetic control MRIs. For one case-control study, we used a generative adversarial model for style-based harmonization to generate site-specific controls. Downstream feature extraction, statistical harmonization and group-level multi-study case-control and case-only analyses were performed twice, using either true or synthetic controls. All effect sizes using synthetic controls overlapped with those based on true study controls. This line of work may facilitate wider inclusion of case-only studies in multi-study consortia.

*Index Terms*— MRI harmonization, Synthetic controls, ComBat, cognitive impairment

## 1. INTRODUCTION

Multi-site data pooling in neuroimaging studies offers significant advantages in terms of sample size and statistical power for generalizability and robustness of findings [1],[2]. Yet, a major challenge in combining imaging data from multiple studies and sites is the variability in image acquisition protocols, scanner hardware, and software. MRI scanner properties such as field strength, manufacturer, gradient nonlinearity, subject positioning, and longitudinal drift have been long understood to increase bias and variance in the measurement of brain volumes [3]. Unfortunately, site or study specific acquisition related variations may be confounded with site-specific population characteristics and study designs (e.g., inclusion/exclusion criteria and regional demographics).

Often, for studies of a particular disease or condition, studies also collect neuroimaging data from demographically matched controls to better isolate condition-specific effects on the brain. Controls are often matched to the clinical population based on several demographic variables, such as ancestry, age, and sex. However, studies that focus exclusively on variations within the clinical populations may not collect controls for case/control comparisons. While this may be adequate for the goals of the original study, this may complicate integration of data with other datasets for future retrospective multi-study or consortium analyses, such as those of the ENIGMA consortium. In these efforts, the first clinical working group collaboration often involves a case/control comparison. This helps ensure feasibility of pooled multi-study analyses, before more condition specific trends, such as medication use or genetic risk, can be evaluated. If a study has not collected controls, in theory, they can only participate in patient-specific analyses.

Many multi-study efforts conduct meta-analyses, where all statistical analyses are conducted exclusively in individual studies and the statistical results are shared and meta-analyzed. However, it is becoming more common to perform centralized analyses where derived data are pooled across studies for a single "mega" analysis. Statistical harmonization is often a first step after pooling subject level imaging features across data collection sites to improve power to detect effects of interest [4]. These can include methods and tools such as those in the ComBat family [5],[6],[7], hierarchical Bayesian regression [8], or brain charts [9],[10]. Across methods, harmonization is often done with respect to controls, and then applied to cases [6],[9],[10]. This can mitigate situations in which patients from different studies are recruited using different criteria, e.g., combining studies focusing on early vs. late disease stages [11], or when there are large differences in sex distributions between one study and others [12]. We emphasize that while features from case-only datasets can be harmonized with those of case-control datasets using ComBat and other methods, this approach is highly susceptible to over-correction, when demographic or clinical information is confounded with acquisition information [13]. In such cases, extra care in both analysis and interpretation of findings are warranted. Matched controls would help ensure that a study can be harmonized and included in large scale consortium initiatives.

To address complications with pooling case-only studies, computational or simulation studies may be used to create synthetic control groups based on existing public, population-based datasets. This involves generating synthetic control data to simulate what might be expected in the absence of the condition under investigation and matching on demographic or clinically relevant distributions of the cases. These matched synthetic controls can then be used as an anchor to harmonize the case-only cohort to other studies for larger-scale statistical analyses.

In this paper, we compare the use of synthetic control T1-weighted MRI (T1w) for pooling case-only MRI datasets with those from other studies for large-scale multi-study analyses. We generated synthetic control T1w for a target dataset by harmonizing the imaging style of control T1w from other public datasets to the imaging style of the AD patients in our target dataset. We used AIBL [14], a dataset with controls and Alzheimer's disease (AD) patients, extracted regional brain volume measures and compared the case-control differences found when using synthetic controls and AD volumes to true controls and AD volumes. Finally, we ran ComBat-GAM [5] to harmonize this dataset with other case/control AD datasets using the 1) true controls and 2) synthetic controls and compared ApoE4 associations within AD participants only.

## 2. METHODS

### 2.1. Datasets

The AD datasets included in this study were the following: Australian Imaging Biomarkers and Lifestyle Study of Ageing (AIBL) [14]; Open Access Series of Imaging Studies-3 (OASIS3) [15]; Alzheimer's Disease Neuroimaging Initiative (ADNI1) [16]; and data from the National Alzheimer's Coordinating Center (NACC).

We created synthetic T1w control data for AIBL and used true AIBL controls for validation. We drew controls from OASIS-Cross-Sectional (OASIS1) [17] and the UK Biobank (UKB) [18] for our synthetic image generation. Demographic, diagnostic, and genetic data as well as T1w MR scans from baseline visits were used (**Table 1**).

### 2.2. Generation of Synthetic Control MR Images

We pooled control T1w MRI from both OASIS1 and the UKB and harmonized their imaging style to the AIBL AD participant imaging style for synthetic generation of AIBL controls.

We used all of the 62 controls from OASIS1 in the age range of our AIBL dementia cases (aged 60-81 years) and randomly selected 210 individuals from the UKB. In UKB, for every age between 60 and 81, we randomly selected five males and five females; for each age-sex combination, there were five individuals selected, two of whom had no ApoE4 risk alleles, two of whom had one, and one of whom had two. All the T1w used for harmonization were bias field corrected, skull stripped using HD-BET [19], registered to MNI template using FSL's *flirt* [20] command with 9 degrees of freedom and then zero padded to obtain image dimension of $256 \times 256 \times 256$.

Site-1 in AIBL had nearly double the number of subjects with AD compared to site-2. Therefore, we opted to split the data from UKB and OASIS in a 2:1 ratio to create synthetic controls for each site from both datasets. In total, 139 out of the 210 T1w images from UKB and 41 images from OASIS1 were harmonized to one AD subject in site-1 in AIBL, and 71 from UKB and 21 from OASIS1 were harmonized to site-2 in AIBL using style transfer harmonization model [21]. Post harmonization steps included removing the zero padding and then moving the image back to subject space from the template space.

**Table 1.** MRI scanner and participant demographic information for the datasets included.

| Dataset | No. of sites | Scanner | Scanner field strength | Age (yrs) | Total/ Male count | CN/ AD count | ApoE4 count 0/1/2 |
|---|---|---|---|---|---|---|---|
| **OASIS1** | 1 | Siemens | 1.5T | 71.40 ±6.05 | 62/ 16 | 62/- | - |
| **UKB** | 1 | Siemens | 3T | 70.62 ±6.17 | 210/ 105 | 210/- | 84/ 84/ 42 |
| **AIBL (Site 1)** | - | Siemens | 3T | 71.84 ±4.99 | 201/ 88 | 178/ 23 | 139/ 54/ 8 |
| **AIBL (Site 2)** | - | Siemens | 1.5T, 3T | 69.63 ±5.45 | 92/ 43 | 83/ 9 | 54/ 35/ 3 |
| **ADNI1** | 47 | Siemens, GE, Philips | 1.5T, 3T | 73.90 ±4.24 | 299/ 145 | 187/ 112 | 165/ 103/ 31 |
| **OASIS3** | 5 | Siemens | 1.5T, 3T | 70.43 ±5.12 | 740/ 316 | 674/ 66 | 470/ 234/ 36 |
| **NACC** | 3 | Siemens, GE, Philips | 1.5T, 3T | 70.94 ±5.98 | 250/ 100 | 203/ 47 | 156/ 79/ 15 |

### 2.3. Image Processing and Feature Extraction

Fast-Surfer [22] was run on final subject space T1w images to extract volumes of subcortical structures (thalamus, hippocampus, amygdala, putamen, caudate, accumbens and pallidum) and ventricles. These metrics were extracted for all the original and synthetic T1w datasets.

### 2.4. ComBat Harmonization of Imaging Features

Volumes derived from T1w scans for all the datasets consisting of multiple sites were harmonized using ComBat Generalized Additive Model (ComBat-GAM; https://github.com/rpomponio/neuroHarmonize) [5].

Two ComBat-GAM harmonization models were trained using site as the batch effect with age, sex and intracranial volume (ICV) covariates; here, age was specified as a nonlinear term using the command *harmonizationLearn*. Models were trained with control volumes from ADNI1, OASIS3, NACC and either true or synthetic control volumes from AIBL. The site-specific parameters derived from these two models were applied to the site-specific AD subsets using the command *harmonizationApply*. In short, the difference between these two ComBat setups was the set of control-derived features (true or synthetic) for AIBL used to train the initial harmonization step.

### 2.5. Statistical Analysis

#### 2.5.1. Case-control differences in AIBL alone

Linear mixed models were run to compare AIBL control and AD subcortical volumes using either true AIBL controls, synthetic AIBL controls, or controls from other cohorts. In all cases, age, sex and intracranial volume (ICV) were added as fixed effects covariates and scanning site as a random effect. The resulting Cohen's *d* effect sizes for true versus synthetic analyses were compared using a paired *t*-test.

#### 2.5.2. Case-control differences in multi-study ComBat harmonized data

Linear regressions were performed to compare control and AD subcortical volumes across ComBat harmonized studies adjusting for age, sex and ICV. This was run once using true AIBL control volumes and once using synthetic control data for comparison.

#### 2.5.3. Case-only regressions in ComBat harmonized data

Another set of linear regressions was performed to calculate associations between ApoE4 count (0/1/2) and harmonized subcortical measures for 1) AD data harmonized using true controls, and 2) AD harmonized using synthetic controls, adjusting for age, sex and ICV. Paired *t*-tests were performed to compare partial correlation (r-value) effect sizes generated when using the true versus synthetic data.

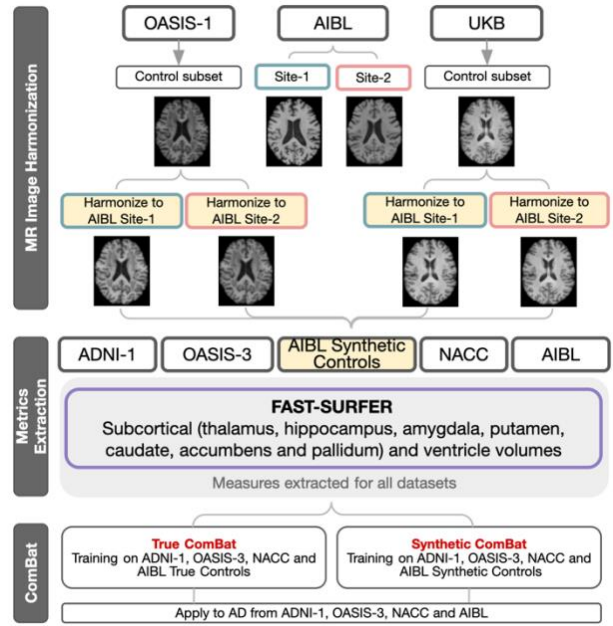

**Figure 1. Flowchart for generating cohort-specific control T1w data.** A subset of control MRI from UKB and OASIS1 were harmonized to two subjects from AIBL (one from each site) to generate synthetic control data. Subcortical and ventricular volumes were extracted for all the datasets using Fast-Surfer. Two ComBat-GAM models were trained on these extracted measures: **True ComBat,** training on controls from ADNI1, OASIS3, NACC and AIBL; **Synthetic ComBat:** training on controls from ADNI1, OASIS3, NACC and Synthetic AIBL. Both of these models were applied to the AD populations of these cohorts for comparison.

## 3. RESULTS

### 3.1. Case-control differences in AIBL alone

Compared to both true and synthetic control participants, AD cases had higher ventricular volumes ($p=4.23 \times 10^{-11}$ for true, $p=1.65 \times 10^{-7}$ for synthetic) and lower bilateral hippocampal volumes ($p=1.10 \times 10^{-13}$ for true, $p=2.64 \times 10^{-10}$ for synthetic). Effect sizes and their confidence intervals for all the extracted measures are shown in **Figure 2**. A paired *t*-test comparing case-control effect sizes across all volumes between the true and the synthetic data was not significant ($p=0.62$) showing that both sets of regressions had similar results. True effects were more likely to be similar to those derived from synthetic images than using the same images in their original (non-harmonized) style, i.e., directly using images from other cohorts.

### 3.2. Case-control differences in multi-study ComBat harmonized data

Paired *t*-test comparing case-control effect sizes across all volumes between the true and the synthetic data after multi-study ComBat harmonization was not significant (*p*=0.94). **Figure 3** shows the case-control effect sizes.

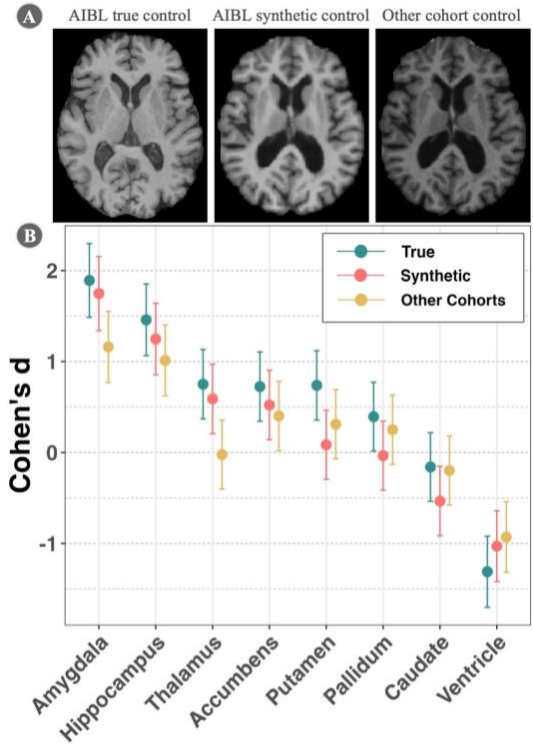

**Figure 2. A**) T1w images of true AIBL control, synthetic AIBL control (OASIS1 harmonized to a subject from AIBL site 1) and unharmonized control from OASIS1 at age 74. **B**) Case-control effect sizes for differences in subcortical and ventricle volumes when comparing AIBL AD patients to either true or synthetic AIBL controls, or when compared to directly using the T1w images from other cohorts without style harmonization.

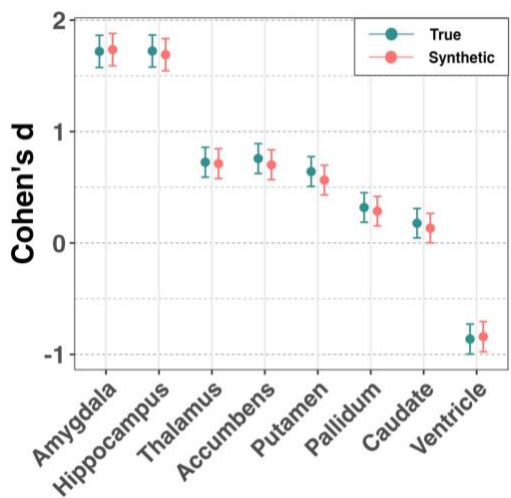

**Figure 3.** Case-control effect sizes with the confidence intervals for all the subcortical and ventricle volumes after ComBat-GAM was run across studies using either true or synthetic AIBL controls.

### 3.3. Case-only regressions in ComBat harmonized data: ApoE4 associations in the AD population

Bilateral amygdala volume was found to be significantly associated with ApoE4 count both when true AIBL controls (r=-0.18; *p*=0.002) and synthetic AIBL controls (r=-0.19; *p*=0.003) were used for ComBat harmonization across studies. **Figure 4** shows a box plot with the residual volumes of the hippocampus and amygdala.

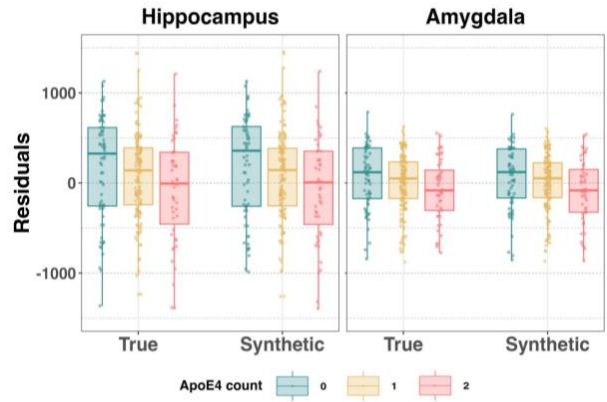

**Figure 4.** Box plot for comparisons of residualized volumes of hippocampus and amygdala for each ApoE4 count using true and synthetic ComBat.

### 4. CONCLUSION AND DISCUSSION

We generated synthetic control T1w images for AIBL by harmonizing the style of control images from UKB and OASIS-1 to the image style of AIBL. We show that a case-only study can be pooled with other case-control studies by using synthetic controls to harmonize data using ComBat-GAM. The effect sizes of AD diagnosis and subcortical volume associations were similar regardless of whether true or synthetic data were used. In AD cases, the associations between ApoE4 count and ComBat harmonized volumes were almost identical when using the true and synthetic control data. We note that this work is preliminary and has several limitations, including only evaluating AD cases where the effects of neurodegeneration are more evident than, for example, those of mood disorders or other psychiatric conditions. While we performed a case-control analysis for validation, the main goal of this work is to better allow for statistical harmonization of case-only datasets for case-only statistical analyses and avoid over-correction. More active research is needed in this field.

### 5. COMPLIANCE WITH ETHICAL STANDARDS

This study was conducted retrospectively using publicly available, anonymized, de-identified human subjects' data. Additional approval was not required beyond that obtained in the original studies.

## 6. ACKNOWLEDGEMENTS

This work is supported in part by NIH grants: R01AG059874, RF1AG057892, P41EB015922, U01AG068057 and R01AG058854. This work was completed using UK Biobank Resource under application number 11559. Acknowledgments for OASIS and NACC can be found at (http://www.oasis-brains.org/#access, https://naccdata.org/publish-project/authors-checklist). Data used in the preparation of this article was obtained from the AIBL funded by the Commonwealth Scientific and Industrial Research Organization (CSIRO). AIBL researchers are listed at www.aibl.csiro.au. Data used in preparing this article were obtained from the ADNI database (adni.loni.usc.edu). As such, many investigators within the ADNI contributed to the design and implementation of ADNI and/or provided data but did not participate in analysis or writing of this report. A complete list of ADNI investigators: http://adni.loni.usc.edu/wp-content/uploads/how_to_apply/ADNI_Acknowledgement_List.pdf.